\documentclass[12pt]{article}
\usepackage{graphicx,epsfig}
\def\lsim{\mathrel{\raise.3ex\hbox{$<$\kern-.75em\lower1ex\hbox{$\sim$}}}}
\def\gsim{\mathrel{\raise.3ex\hbox{$>$\kern-.75em\lower1ex\hbox{$\sim$}}}}
\newcommand{ \slashchar }[1]{\setbox0=\hbox{$#1$}   
   \dimen0=\wd0                                     
   \setbox1=\hbox{/} \dimen1=\wd1                   
   \ifdim\dimen0>\dimen1                            
      \rlap{\hbox to \dimen0{\hfil/\hfil}}          
      #1                                            
   \else                                            
      \rlap{\hbox to \dimen1{\hfil$#1$\hfil}}       
      /                                             
   \fi}                                             %

\def\be{\begin{equation}}
\def\ee{\end{equation}}
\def\bea{\begin{eqnarray}}
\def\eea{\end{eqnarray}}
\def\bec{\begin{center}}
\def\eec{\end{center}}
\def\atversim#1#2{\lower0.7ex\vbox{\baselineskip\zatskip\lineskip\zatskip
  \lineskiplimit 0pt\ialign{$\matth#1\hfil##\hfil$\crcr#2\crcr\sim\crcr}}}

\renewcommand{\thefootnote}{\fnsymbol{footnote}}

\newcounter{appendixc}
\newcounter{subappendixc}[appendixc]
\newcounter{subsubappendixc}[subappendixc]

\renewcommand{\appendix}[1] {\vspace*{0.6cm}
        \refstepcounter{appendixc}
        \setcounter{figure}{0}
        \setcounter{table}{0}
        \setcounter{equation}{0}
        \renewcommand{\thefigure}{\Alph{appendixc}.\arabic{figure}}
        \renewcommand{\thetable}{\Alph{appendixc}.\arabic{table}}
        \renewcommand{\theappendixc}{\Alph{appendixc}}
        \renewcommand{\theequation}{\Alph{appendixc}.\arabic{equation}}
        \noindent{\bf Appendix \theappendixc #1}\par\vspace*{0.4cm}}

\begin{document}
\begin{titlepage}
\rightline{\vbox{\halign{&#\hfil\cr \cr
\cr }}} \vskip .5in
\begin{center}

{\Large\bf A Uniform Description of the States Recently Observed at
B-factories}
\begin{figure}
{\hspace{12.7cm}\vbox{\halign{&#\cr &GUCAS-CPS-07-006 \cr}}}
\end{figure}
\vskip .5in \normalsize {\bf Cong-Feng Qiao
\footnote{Email: qiaocf@gucas.ac.cn}}\\
\vskip .5cm

Dept. of Physics, Graduate University, the Chinese
Academy of Sciences\\
YuQuan Road 19A, Beijing 100049, China\\[2mm]
Theoretical Physics Center for Science Facilities
(TPCSF), CAS\\
YuQuan Road 19B, Beijing 100049, China\vskip 2.3cm

\end{center}

\begin{abstract}
\normalsize

The newly found states Y(4260), Y(4361), Y(4664) and Z$^\pm$(4430) 
stir broad interest in the study of spectroscopy in a typical
charmonium scale. The Y(4260) which was observed earlier has been
interpreted as hybrid, molecular state, and baryonium, etc. In this
note we show for the first time that these new structures, which are
hard to be interpreted as charmonium states, can be systematically
embedded into an extended baryonium picture. According to this
assignment, the so far known characters of these states are
understandable. And, in the same framework, we make some predictions
for experimenters to measure in the future.

\end{abstract}

\vspace{1cm} \hspace{0.3cm} PACS number(s): 12.39.Mk, 14.20.Lq,
14.20.Pt

\renewcommand{\thefootnote}{\arabic{footnote}}
\end{titlepage}

Recently, a series of new resonances in the range of 3.8 $\sim$ 5.5
GeV were observed in experiment. Among them, the Y(4260), Y(4361),
Y(4664) and Z$^\pm$(4430) have distinct characters from others. They
are hard to be embedded into the regular charmonium spectroscopy,
which means that they may have exotic natures. Of these four states,
Y(4260) was first observed by the BaBar Collaboration in the
initial-state radiation process $e^+ e^- \rightarrow \gamma_{ISR}\
\pi^+ \pi^- J/\psi$ with a mass of 4259 $\pm 8 \pm 4 $ MeV and total
decay width of 88 $\pm 23 \pm 5 $ MeV \cite{babar1}. Obviously, this
state possesses the same quantum numbers as photon, i.e. $J^{PC} =
1^{--}$. Since the mass of Y(4260) lies in the range of regular
charmonium states, a natural interpretation for it is that it
contains charm-anticharm constituents. However, although the
structure exhibits at about $4.26$ GeV, much higher than the
open-charm threshold, so far the exclusive $D \bar{D}$ decay mode
has not been reported. The Y(4260) was later confirmed by CLEO
\cite{cleo} and Belle Collaborations \cite{belle1}.

In theory, several models were proposed to explain it. Except for
Ref.\cite{felipe} where the Y(4260) was still interpreted as a
normal member in the charmonium spectra, a common belief is that it
is rather an exotic (or cryptoexotic) state. The authors of Refs.
\cite{zhu,kou,close} proposed that the new state is a charmonium
hybrid that is constructed by a pair of charm-anticharm quarks and a
gluon. Ref. \cite{maiani} treats Y(4260) as the first orbital
excitation of a diquark-antidiquark state $[cs][\bar{c}\bar{s}]$. In
Ref. \cite{li}, Liu {\it et al.} explained the resonance as a
molecular state being composed of $\rho$ and $\chi_c$, while the
authors of Ref. \cite{yuan} took it as a $\omega$ and $\chi_c$
molecular state. And, in Ref. \cite{qiao} the Y(4260) was
interpreted as a baryonium state containing hidden charm and is made
out of $\Lambda_c^+$-$\Lambda_c^-$, or triquark-antitriquark pair.

Of the existing models, the CLEO data are still consistent with the
hybrid and baryonium models. And, it should be noticed that there
exists a misunderstanding that the CLEO measurement of the ratio
\cite{cleo}
\begin{eqnarray}
\Gamma[Y(4260)\rightarrow  \pi^0\pi^0\psi] \approx
\frac{1}{2}\Gamma[Y(4260)\rightarrow  \pi^+\pi^-\psi]\;
\end{eqnarray}
disfavors the baryonium model. In fact, it is exactly the baryonium
model prediction \cite{qiao}, and in principe it should also comply
with other models having the $J/\psi \pi \pi$ decay mode.

The Y(4361) was first observed by Babar Collaboration in the
subsequent search for Y(4260) in process of $e^+ e^- \rightarrow
\gamma_{ISR}\ \pi^+ \pi^- \psi(2S)$ \cite{babar2}. They found no
obvious signature of Y(4260), but rather a new structure at $4324
\pm 24$ MeV with a width of $172 \pm 33$ MeV. Recently, this
observation is confirmed by Belle Collaboration, and they find that
in fact the structure 4324 MeV is composed of two resonant
structures \cite{belle2}. One at M = $4361 \pm 9 \pm 9$ MeV with a
width of $74 \pm 15 \pm 10$ MeV, another at M = $4664 \pm 11 \pm 5$
MeV with a width of $48 \pm 15 \pm 3$ MeV, which we will refer as
Y(4361) and Y(4664) respectively in the following discussion.

Although it is a bid hard to imagine that there are many $1^{--}$
vector hybrids in-between the range of one GeV, to distinguish the
quark-antiquark-gluon hybrid model from the multiquark models in a
single neutral state measurement looks difficult, since the gluon of
the hybrid can always split into quark pairs. A characteristic that
can distinguish multiquark states from hybrids or charmonia is of
the non-zero charge of the states in the charmonium energy region. A
new structure of Z(4433) was reported found in the process of $B
\rightarrow K Z(4433) \rightarrow K \pi \psi(2S)$ by Belle
Collaboration, in the recent Lepton-Photon 2007 symposium
\cite{belle3}. It has a mass of $M = 4433\pm 4$(stat)$\pm 1$(syst)
MeV and width $\Gamma = 44^{+17}_{-13}$(stat)$^{+30}_{-11}$(syst)
MeV. One unique nature of this new state is that it possesses
electric charge(the $Z^+$ and its charge conjugation $Z^-$ are both
observed). Since the statistical significance of the observation is
greater than $7\sigma$, it deserves to be treated seriously.

Soon after the announcement of the observation of this new
structure, Z(4433), there appear couple of theoretical speculations
on its nature. Refs. \cite{ros} and \cite{zhao1} suggest that it
might be a $D^*(2010)-\bar{D}(2420)$ rescattering resonance, because
its mass is in proximity to the $D^*(2010) \bar{D}(2420)$ threshold,
while Bugg \cite{bugg} takes it as a $D^*(2010) \bar{D}(2420)$
threshold cusp. Maiani {\it et al}. \cite{maiani1} interpret it as a
tetraquark bound state, etc.

Since there are increasingly numerous new structures observed
recently in the charmonium sector, it is tempting to give a uniform
description of them, or at least some of them. According to
Ref.\cite{qiao}, the Y(4260) can be interpreted as a baryonium state
of $\Lambda_c-\bar{\Lambda}_c$.  If we extend this baryonium picture
to include $\Sigma_c^0$ as a basic ingredient, the $\Lambda_c$ and
$\Sigma_c^0$ can be taken as basis vectors in two-dimensional space,
which is similar to the proton and neutron in constructing the pion
by Fermi and Yang more than fifty years ago \cite{fermi}.
Approximately, we assume that the transformation in this
two-dimensional "C-spin" space is invariant, i.e., there exists a
SU(2) symmetry between $\Lambda_c$ and $\Sigma_c^0$. Then, from this
doublet one can make up four baryon-antibaryon configurations, the
"C-spin" triplet and singlet(note that the charged $\Sigma_c$ and
$\Lambda_c$ can give a similar mass degenerate structure). That is:
\begin{eqnarray}
B^+_1\equiv |\Lambda_c^+ \; \bar{\Sigma}_c^0>~~~~~~~~~\nonumber\\
{\rm Triplet:}\;\;\;\;\; B^0_1\equiv \frac{1}{\sqrt{2}}(|\Lambda_c^+
\;\bar{\Lambda}_c>\; -\; |{\Sigma}_c^0 \bar{\Sigma}_c^0>)\\
B^-_1\equiv |\Lambda^-_c\; {\Sigma}_c^0>~~~~~~~~~\nonumber
 \label{triplet}
\end{eqnarray}
and
\begin{eqnarray}
{\rm Singlet:}\;\;\;\;\; B^0_0\equiv \frac{1}{\sqrt{2}}(|\Lambda_c^+
\;\bar{\Lambda}_c>\; + \; |{\Sigma}_c^0 \bar{\Sigma}_c^0>)\ .
\label{sin}
\end{eqnarray}
This is just an imitation of isospin for proton and neutron system.

Although in the baryonium picture there surely exits large symmetry
broken effects, it is still possible to make a qualitative
description of the mass spectrum. For $Z^{+}(4433)$ and
$Z^{-}(4433)$, which have the quantum number $J^{P} = 1^+$ (however,
the $0^-$ and $1^-$ are not excluded \cite{belle3}) and isospin I =
1, we find their correspondences are $\Lambda_c^+ \;
\bar{\Sigma}_c^0$ and $\Lambda_c^- \; {\Sigma}_c^0$, respectively.
Since the $Z^{\pm}(4433)$ are found in the $ K \pi \psi(2S)$ channel
and there is no signature found at the same energy region in the $K
\pi J/\psi$ mode \cite{steve}, it is plausible to consider
$Z^{\pm}(4433)$ to be the first radial excitation of $\Lambda_c \
{\Sigma}_c^0$, that is $B_1^{\pm*}$. Otherwise, on the other hand,
the mass difference between Y(4260) and Z(4433) is too big, provided
they are in the same triplet in (2). Therefore, in this assignment
there should exist the charged states at mass of 4330 MeV or so,
which correspond to the ground states composed of $\Lambda_c^{\pm}
\; {\Sigma}_c^0$, i.e. $B^\pm_1$. The magnitude of 4330 MeV is
obtained from the following arguments.

We know that Y(4361) and  Y(4664) were observed in the $e^+ e^-
\rightarrow \gamma_{ISR}\ \pi^+ \pi^- \psi(2S)$ process
\cite{belle2}, and Belle found no structures in the same mass region
in process of $e^+ e^- \rightarrow \gamma_{ISR}\ \pi^+ \pi^-
J/\psi$. One of the possibilities is that these states are the first
radial excitations of the heavy hidden quark pair, which may easily
decay to $\psi(2S)$ rather than to $J/\psi$, relatively. Explicitly,
Y(4361) might be the radial excited state of Y(4260) and Y(4664) the
radial excited state of the C-spin singlet (\ref{sin}). Of course,
the reason why the energy gap between ground state and the first
radial excitation is so small needs further investigation. From the
formerly defined C-spin symmetry, the ground state of Z(4433) should
be around 4330 MeV, and the ground state for Y(4664) should be
around 4560 MeV. In fact from Ref.\cite{belle2} one can read an
unclear structure around the 4500 MeV, which is left for further
confirmation.

Except for above arguments, the PDG data read \cite{pdg}
\begin{eqnarray}
M(Y(4361)) - 2 M(\Lambda_c) \approx M(Y(4664)) - 2 M(\Sigma_c)
\approx M(Z(4433)) - M(\Lambda_c) - M(\Sigma_c)\ .
\end{eqnarray}
Due to the existing uncertainties in the experimental measurement,
the above relations should only be treated as an order-of-magnitude
estimation. Nevertheless, the equation (4) in some sense supports
the baryonium picture.

\begin{table}[htbp]
\caption{The experimental measurements and baryonium model
predictions (speculations) for spin triplet states $B$ and their
radial excitations $B^*$. The question mark means unobserved in
experiment, and the pseudoscalar partners of the listed states are
all missing. The numbers in the brackets are in the units of MeV.}
\label{Table-spectrum}
\begin{center}
\begin{tabular}{cccc}
\hline \hline

$B_1^+(4330?)$ &\qquad $B_1^0(4260)$ &\qquad $B_1^-(4330?)$ &\qquad
$B_0^0(4560?)$
\\
\hline $B_1^{+*}(4430)$ &\qquad $B_1^{0*}(4361)$ &\qquad
$B_1^{-*}(4430)$ &\qquad $B_0^{0*}(4664)$
\\
\hline \hline
\end{tabular}
\end{center}
\end{table}

Different from the model in Ref.\cite{qiao}, where the Y(4260) was
treated as a $\Lambda_c\  \bar{\Lambda}_c$ bound state, in the
extended picture, the physical states of Y(4260) and Y(4560)(suppose
the latter exists) are the mass eigenstates of two orthogonal
vectors of (2) and (3) as given in Table 1. From the experimental
measurement, one can infer that the Y(4260) has a large component in
$\Lambda_c\ \bar{\Lambda}_c$ configuration and Y(4560) is mainly in
$\Sigma_c^0\ \bar{\Sigma}_c^0$ configuration. If necessary, the
mixing angle can be straightforwardly calculated.

In summary, in this paper we try in the first time to give a
systematic description (speculation) of the newly observed
structures(states) in B-factories. We assume that these states can
be embedded into the multiquark picture, i.e., $(c[n][n] +
\bar{c}[\bar{n}][\bar{n}])$, where the [n] denotes for the light
quarks u and d; and the hexaquark states are configured by the
triquark-antitriquark clusters, namely, the baryonia. In comparison
with the Teraquark model, the hexaquark scheme predicts that more
rich spectrum in the charmonium sector may exist.

In this extended baryonium picture, there is an approximate C-spin
symmetry. Four classes of baryonium states are predicted, three in
triplet and one in singlet. This model is partially supported by the
experimental data. We predict two vector-like structures could exist
around 4560 MeV and 4330 MeV. The former should be neutral and the
latter is charged. In addition, since the concerned states, Y(4260),
Y(4361), Y(4664) and Z$^\pm$(4430), are vector states(the last one
is not fully confirmed), if the baryonium picture is correct, their
pseudoscalar para-baryonium partners should also exist. All these
wait for future experiment to examine.

Last, it should be noted that our interpretation for the newly
observed states Y(4260), Y(4361), Y(4664) and Z$^\pm$(4430) as
baryonia is merely based on the quark model. From the success of
quark model in describing the normal meson and baryon, we expect
that this picture may give a qualitative description of those newly
observed states at relatively low energy. For fine structures and
exquisite natures of them, the involvement of QCD dynamics is
unavoidable, which, although is one of our aims for further
investigation, is beyond the scope of this brief report.
\vspace{1.3cm}
\par
\par
{\bf Acknowledgments} \vspace{.2cm}

This work is supported in part by the National Natural Science
Foundation of China and by the Scientific Research Fund of GUCAS
(NO. 055101BM03).

\end{document}